\documentclass[conference]{IEEEtran}
\IEEEoverridecommandlockouts
\usepackage{cite}
\usepackage{amsmath,amssymb,amsfonts}
\usepackage{graphicx}
\usepackage{textcomp}
\usepackage{xcolor}

\usepackage{multirow}
\usepackage{framed,enumitem}
\usepackage{soul}
\usepackage{subfig}
\usepackage{tikz}
\usepackage{tcolorbox}
\usepackage{algorithm}
\usepackage{algpseudocode}
\usepackage{url}
\usepackage[hidelinks]{hyperref}
\usepackage{amsthm}
\newtheorem{definition}{Definition}
\usepackage{bm}

\def\BibTeX{{\rm B\kern-.05em{\sc i\kern-.025em b}\kern-.08em
    T\kern-.1667em\lower.7ex\hbox{E}\kern-.125emX}}
\begin{document}

\title{Reasoning-Based Software Testing}

\author{\IEEEauthorblockN{Luca Giamattei, Roberto Pietrantuono, Stefano Russo}
\vspace{6pt}
\IEEEauthorblockA{\textit{DIETI, Universit\`a degli Studi di Napoli Federico II}\\
Via Claudio 21, 80125 - Napoli, Italy \\
\{luca.giamattei, roberto.pietrantuono, stefano.russo\}@unina.it}
}

\maketitle

\begin{abstract}
With software systems becoming increasingly pervasive and autonomous, our ability to test for their quality is severely challenged. %
Many 
systems are called to operate in uncertain and highly-changing environment, not rarely required to make %
intelligent decisions by themselves. This easily results in an intractable state space to explore at testing time. 
The state-of-the-art 
techniques try to keep the pace, e.g., by augmenting the tester's intuition with 
some form of (explicit or implicit) learning from 
observations to 
search this space 
efficiently. For instance, they exploit 
historical data to drive the search (e.g., ML-driven testing) or the tests execution data itself (e.g., adaptive 
or search-based testing).  
Despite the indubitable advances, the need for smartening the search in such a huge space keeps to be pressing.

We introduce Reasoning-Based Software Testing (RBST), a new way of thinking at the testing problem as a causal reasoning task. Compared to mere intuition-based or state-of-the-art learning-based strategies, we claim that causal reasoning more naturally emulates the process that a human would do to ``smartly'' search the space. RBST aims to mimic and amplify, with the power of computation, this ability. The conceptual leap can pave the ground to a new trend of techniques, which can be variously instantiated from the proposed framework, by exploiting the numerous tools for causal discovery and inference. Preliminary results reported in this paper are promising.

\end{abstract}

\begin{IEEEkeywords}
Causal reasoning, Software Testing
\end{IEEEkeywords}

\section{Introduction}

Software testing is ultimately about \textit{prediction}. In deriving tests, a tester basically asks her/himself: \textit{What input makes the system fail?} 
S/he tries to envisage how the system would behave for some specific input in a given execution scenario. 
To get support for this prediction, s/he usually relies on some auxiliary information to derive failure-exposing tests (e.g., using corner/adversarial inputs, coverage measures, discrepancy/similarity measures with respect to an observed context). %
Existing techniques exploit the tester's belief about what is expected to correlate with failures, to automate testing tasks. In a learning-supported strategy, the belief 
is corroborated and complemented by past observations (testing/operational data) for training a Machine Learning (ML) model 
\cite{Abdessalem18a,Durelli19,Pang13} or, unsupervisedly, exploiting the feedback from tests execution
to derive better tests in the next step 
(e.g., adaptive testing \cite{Mullins18}, search-based software testing  
\cite{McMinn11,Harman15}). %
Learning 
supports various 
tasks, such as 
tests generation \cite{Cegin20,Li19nier}, test selection and prioritization 
\cite{Grano18,Busjaeger16}, test execution \cite{Stocco20,Stocco21}.

This current approach to testing has inherent limitations: it either relies only on intuition (expertise/experience) to ``guess'' the right failure-correlated information, then devising a technique from it, or is focused on learning failing patterns from past observations to predict the best next tests, 
but assuming that the future context resembles the past. 

While human reasoning is, in principle, great at ``guessing'' failure-related information, it, clearly,  cannot scale.  %
Learning from the past 
indeed helps navigate the search space, but it is just a palliative: %
correlations learnt on observations in a certain context are exploitable to make ``predictions'' solely based on what seen \cite{Pearl18}. 
This hides a conceptual glitch: 
with current learning-based 
techniques such as ML-driven testing, we are implicitly turning 
the \textit{What makes it fail?} question  into: \textit{What is more correlated to failure?} But this is not the kind of prediction a tester tries to do: with that question, s/he means: \textit{What input \underline{causes} the system to fail?} 

The former question reads as: \textit{Given the same context in which I learned the model (i.e., the same data distribution), what  output $Y$ do we expect if the input $X$ \underline{happens to be} equal to $x$?} %
On this basis, for instance, we select or prioritize tests that are more similar to failure-causing tests observed in the past, and ML is great at this task. %

The causal question reads as: \textit{What output $Y$ do we expect if the input $X$ \underline{is actively set} to $x$} (hence, if we change X's distribution)? The tester's question is inherently \textit{causal}, but 
a learning-from-association paradigm 
like ML %
cannot handle causality \cite{pearl2009}. It indeed
amplifies our pattern search ability, which is a great added-value; however, we can do much more, and rather amplify our causal reasoning ability. The latter is not limited to learning from observations, but it learns from hypothesizing  \textit{interventions} on variables of interest. %

This work proposes a conceptual leap in the way machine and human cooperate to intelligently explore the huge search space and derive tests. Machine should support and boost human reasoning far beyond the mere search for patterns in past observations. 
We claim that automated \textit{causal reasoning} is the next step we should take in software testing, as it gives the ability to infer knowledge proactively, and to ask \textit{what happens if} questions in a world different from the one observed. %

Hereafter, we present our proposal to cast the testing problem as a  causal-reasoning task, called \textit{Reasoning-Based Software Testing} (RBST). 
We then preliminarily evaluate a basic instance of RBST, for testing an Autonomous Driving System against adaptive testing and an ML-driven search-based technique. Results show the benefit of exploiting cause-effect relations to derive safety-violating tests.

\label{sec:intro}

\section{Background}

\subsection{Causal Inference}

Causality is the influence by which an event contributes to the production of other events \cite{wire}. For decades statisticians have tried to explain causality through 
statistical methods identifying associations between variables (e.g., correlation, regression), which however  
cannot distinguish  
between cause and effect. 
These are limited to what Pearl called the first rung of the ladder of causation\cite{Pearl18}, that is ``Association". 
Causal reasoning allows to stair up to the second and third rung, respectively, learning by ``\textit{doing}'' (interventions) and by ``\textit{imagining}'' (counterfactuals). 
Research on causality 
has two main branches, causal inference, aimed at quantifying the effect of changing one or more 
variables on an outcome of interest, and causal discovery, aimed at extracting a causal model from observational data. A causal model is a mathematical representation of causal relationships between variables. A common type of model is defined as follows: 

\begin{definition}
(Structural Causal Model) An SCM is a Directed Acyclic Graph $\mathcal{G}=(\bm{X}, \mathcal{E})$, where nodes $\in \bm{X}$ are random variables and edges $\in \mathcal{E}$ are the causal relationships between them, associated with a collection of structural assignments $X_k:=f_k(Pa(X_k), U_k)$ that define the (endogenous) random variables $X_k$, as function of their parents $Pa(X_k)$ and of (exogenous) independent random noise variables $U_k$. %
\end{definition}

In causal inference, we are interested in the distribution of an outcome variable $X_k$ after setting a  variable $W$ to a certain value $w$ (i.e., \textit{doing} an intervention), rather than after just \textit{seeing} an occurrence $W=w$ (i.e., $P(X_k|W=w)$) like in ML.  %
Pearl introduced the  \textit{do-operator}, a mathematical representation of physical intervention, written as $P(X_k|do(W_i=w))$ \cite{Pearl09}.  
An intervention $do(W=w)$  changes the SCM graph (namely, it modifies the distribution), by  removing the causal relations with its predecessors (i.e.,  deleting the $Pa(W)\rightarrow W$ arrows). Thus:  $P(Y|do(X_i=x)) \neq P(Y|(X_i=x))$. %
\begin{definition}
(Intervention distribution) The probability $P(X_k|do(W=w))$ over an SCM is the distribution entailed by the SCM obtained by replacing 
the definition $X_k:=f_k(Pa(X_k), U_k)$ with $X_k:=w_k$. 
\end{definition}

\begin{figure}[b]
	\centering
	\subfloat[Original SCM]{
	 	\includegraphics[width=0.30\columnwidth]{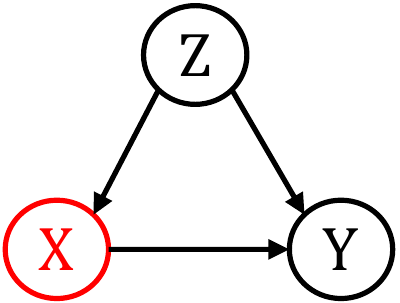}
	 	\label{original:scm}} 
	\subfloat[SCM after intervention]{
	 	\includegraphics[width=0.30\columnwidth]{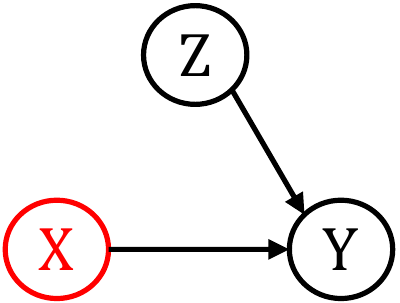}
            \label{modified:scm}}
   \caption{\protect\subref{original:scm} A sample Structural Causal Model; \protect\subref{modified:scm} Effect of an intervention on \textit{X}}
\label{fig:example}
\end{figure}

Figure \ref{fig:example}\subref{original:scm} shows a partial SCM , where \textit{X} causally affects \textit{Y}, and \textit{Z} affects both \textit{X} and \textit{Y}. Considering \textit{X} as intervention variable $W$, an intervention changes the graph (hence the distribution), by removing causal relations with its predecessors, as shown in Figure \ref{fig:example}\subref{modified:scm}.

From an intervention, it is possible to estimate causal effects through the \textit{do-calculus}\cite{pearl2009}: it allows expressing a \textit{do} operation in terms of conditional distributions of a set of related variables, properly identified by graph patterns (e.g.,  \textit{back-door, front-door, instrumental variable}). %
The effect can be quantified by various metrics, the most common one being the \textit{average treatment effect} (ATE)\footnote{With $W$ binary,  %
ATE $= E[X_k|do(W=1)]- E[X_k|do(W=0)]$ \cite{wire}.}.

An additional opportunity with causal inference is to use  counterfactuals. 
A counterfactual is a proposition in the form of a subjunctive conditional such as ``if $W$ had been $w$, then $X_k$ would have been $x_k$". 
This interestingly allows answering questions like \textit{What would have happened if}, thus enabling the exploration of an alternative hypothetical past. 

In software engineering, causality is partly used in a few studies, e.g., to support root cause analysis and diagnosis \cite{Wu21}, debugging\cite{Johnson20}, fault localization \cite{Baah11,kucuk21}, and for interpretability of machine learning models  \cite{Sun22,Chattopadhyay19}.

\subsection{Causal Discovery}
\label{subsec:discovery}
In causal inference, the causal structure is often assumed. There are three main alternatives to build a causal model. The first one consists in intervening on variables and observing the post-intervention probability distributions 
(i.e., controlled experiments). This can also be done with \textit{soft} interventions, which influence the intervened variables distribution without setting it to a fixed value \cite{Eberhardt07}, \cite{Kocaoglu2019}.

The second option consists in using causal discovery algorithms, which extract a causal structure from observational data, hence avoiding expensive (or even technically infeasible) controlled experiments. The last decades have seen advances in the development of such algorithms to enable better use of ``big data". Causal discovery algorithms aim to seize the causal structure from observational data taking statistical dependencies (or independencies) as indicators of causal relations (or lack thereof) \cite{Eberhardt2017}.
Causal discovery algorithms can be divided into constraint-based (e.g., PC, FCI, RFCI), score-based (e.g., GES, FGES, GFCI), and FCM-based (e.g., LinGAM)\cite{Glymour19}.

Constraint-based algorithms use independence tests on observed data in order to determine a set of edge constraints\cite{Spirtes21}. Although these algorithms have the benefit of being broadly applicable, they might not perform well without large sample sizes \cite{Glymour19}.
Score-based algorithms use adjustment criteria like the Bayesian Information Criterion to maximize the score given to candidate graphs. The goodness-of-fit tests are used in place of the conditional independence tests.
Finally, using various model assumptions (e.g., linear non-Gaussian parameterization for LinGAM), FCM-based algorithms try to identify the true causal structure by determining the causal direction of edges.

In addition, due to its interpretable nature, a causal model could also be manually built or refined with domain experts' knowledge.

 \begin{figure*}[t]
	\centering
     \includegraphics[width=0.6\textwidth]
    {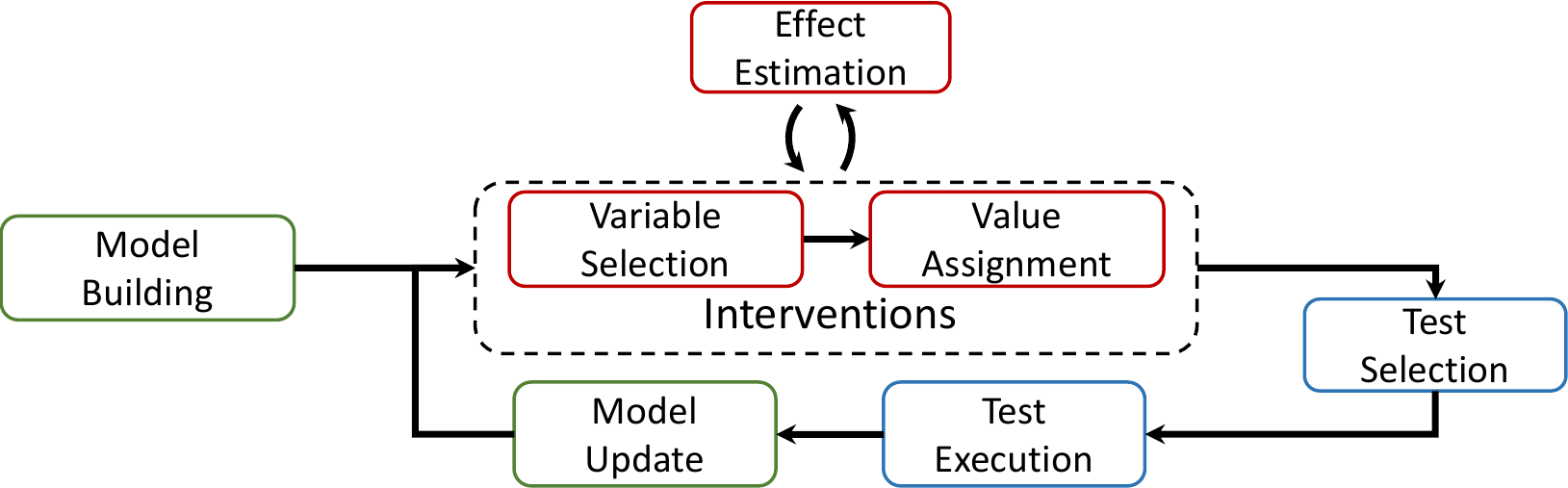}
    \caption{Reasoning-Based Software Testing}
    \label{fig:flow} 
 \end{figure*}
\label{sec:SBST}

\section{Reasoning-Based Software Testing}

Causal inference aims at  estimating the causal effect of one or more variables (treatment) over a certain outcome of interest \cite{wire}. 
In software testing, 
a direct application is for test data generation, wherein the aim is to identify input combinations that maximize/minimize the test output (e.g., performance metrics or safety-related variables) or a  metric of interest (e.g, coverage). 
Reformulating testing objectives as causal questions is possible given the underlying causal structure of the system under test. For instance, in the context of test data generation, cause-effect relations among variables give the possibility to query the model with questions such as: \textit{``What happens to the test output Y if we fix the test input $X=x$?"}. If the model's predictions are accurate, the test space can be explored by using the model, without executing any tests, so as to select and actually run only the most promising tests. %

Test case generation is the focus of this paper (supported by experimental results). However, causal reasoning can assist testers in many other tasks such as regression test selection and prioritization, test suite minimization, and test planning, or even to analyse results of tests execution (and support debugging) 
with the use of counterfactuals (i.e.: as \textit{Would have we still observed the failure if we had fixed $X=x$?}) %

\subsection{Test Case Generation}
Efficient test case generation requires the ability of searching the input space intelligently, so as to identify combinations satisfying the testing goal (e.g., maximize fault detection or coverage) with reasonable cost.  
This exploration process is 
redefined as a causal reasoning task. The envisaged Reasoning-Based Software Testing process is depicted in Figure \ref{fig:flow}). 

There is a wide range of possibilities to instantiate the process. Possible alternatives for the key steps of RBST are 
listed in 
Table \ref{tab:operators}, with a non-exhaustive list of alternatives. The selection of a combination of alternatives forms a test generation strategy.

\begin{table}[hb!]
\centering
\vspace{-2pt}
\caption{Alternatives in instantiating the RBST process}
\label{tab:operators}
\vspace{-2pt}
\begin{tabular}{l|l|l}
\textbf{Step} & \textbf{Description} & \textbf{Alternatives} \\ \hline
\textit{\begin{tabular}[c]{@{}l@{}}Model \\ building\end{tabular}} & \begin{tabular}[c]{@{}l@{}}How to build the \\ causal model\end{tabular} & \begin{tabular}[c]{@{}l@{}}Causal Discovery \\ Domain Expert \\ Controlled Experiments\end{tabular} \\ \hline
\textit{\begin{tabular}[c]{@{}l@{}}Intervention \\variable\\ selection\end{tabular}} & \begin{tabular}[c]{@{}l@{}}How to select the \\ variable(s) for the \\ intervention\end{tabular} & \begin{tabular}[c]{@{}l@{}}
\textit{\underline{Criteria:}}\\ Uncertainty, Confidence\\
Test objective(s)\\ 
Diversity%
\end{tabular} \\ \cline{1-2}
\textit{\begin{tabular}[c]{@{}l@{}}Intervention \\value\\ assignment\end{tabular}} & \begin{tabular}[c]{@{}l@{}}How to set the \\ intervention value \\ for the selected \\ variable(s)\end{tabular} & \begin{tabular}[c]{@{}l@{}}\textit{\underline{Strategies:}}\\
Sampling\\% (adaptive, \\ random) \\
Search-based \\ Adaptive\\
Learning-based\\ %
Exhaustive\end{tabular} \\ \hline
\textit{Effect estimation} & \begin{tabular}[c]{@{}l@{}}How to estimate \\ the effects  of an \\intervention\end{tabular} & \begin{tabular}[c]{@{}l@{}}Analytic\\ Simulation-based\end{tabular} \\ \hline
\end{tabular}
\end{table}

The causal model encodes the causal structure knowledge enabling the inference. 
It is built initially and iteratively refined as more data becomes available. As mentioned in Section \ref{subsec:discovery}, controlled (or soft) experiments, Causal Structure Discovery (CSD) from data (such as past executions of an initial bunch of tests), and, non-alternatively, using 
domain knowledge are the options.  In the latter case, it is worth to stress that the validity of the model is then cross-checked with data; should data not support an assumed relation, the domain expert can refine the model for a next iteration.

The model can be queried via a set of \textit{interventions}. Each intervention  allows estimating the effect of a potential change analytically or via simulation. Thus, an intervention produces a \textit{hypothetical} 
test case, namely a hypothesis for a test along with the expected effect if such test would be executed.  %
The RBST user is here required to develop a strategy for selecting the variable(s) on which to intervene, and the values to assign. An option is to 
select the intervention maximizing the information gained (i.e., minimize the uncertainty) about the true graph  
-- hence intervene to better learn the graph  \cite{Steyvers03}. 

On the other hand, testers might want to set an intervention maximizing/minimizing the desired test objective(s) (e.g., increase coverage, produce critical outputs)
or maximize diversity (e.g., impacting  more effects together, helpful in multi/many-objective testing). A combination thereof can  be set up, e.g., to trade model accuracy and  reward: initially, uncertainty-driven interventions improving the model could be better, gradually replaced by objective-driven intervention. 

Regardless of the chosen criterion, both the variable and value selection/assignment can be implemented in several ways, such as: 
via probabilistic sampling (e.g., (non-)uniform random sampling), search-based or learning-based techniques, adaptive strategies (i.e., using previous selections to drive the next ones), or even exhaustively (depending on the context, the intervention computation time could be negligible compared to real tests execution time). 
 
Each intervention produces an effect, typically quantified by the Average Treatment Effect (ATE), although other metrics can be of interest (e.g.: ATE on Treated (ATT), Conditional ATE (CATE)) \cite{wire}. The effect estimate is usually obtained analytically, by using libraries such as \texttt{DoWhy} \cite{Sharma21}, or via simulation, namely by sampling from the post-intervention distributions and  computing the desired effect estimate. 

One (or more) actual tests are selected from the so-obtained set of \textit{hypothetical} tests. 
The basic choice is to get maximum-effect test(s), but alternatives are worth to be explored, e.g., to improve diversity. 
The tests are then executed; this also enriches the knowledge to update the model. 
Finally, the updated knowledge can be used to double check if the estimated effect is significant or not (i.e., if are due to chance), via refutation tests and confidence interval computation \cite{Sharma21}.

\label{sec:CRBST}

\section{Evaluation}
\subsection{Context}

We evaluate a basic instance of RBST in the context of Autonomous Driving System (ADS) testing for critical scenarios generation. We use Pylot\cite{gog2021pylot} as  ADS, and CARLA as simulator \cite{Dosovitskiy17}. 

A test scenario\footnote{Details are in the replication package available at: \url{https://github.com/uDEVOPS2020/Replication-package-Reasoning-Based-Software-Testing}.} is defined by:
\begin{itemize}
\item sixteen (categorical) variables, including, among others, road type, presence of cars, weather conditions; 
\item one output, namely the minimum distance $d$ from other vehicles.
\end{itemize} 
The objective of a testing session is to find safety violations, namely scenarios in which the event $d=0$ (collision) occurs at least once. 

We run the experiments on a virtual machine deployed on the Google Cloud Compute Engine Platform,\footnote{\url{https://cloud.google.com/compute.}} configured with Ubuntu version 18.04. The libraries used for causal discovery and inference are, respectively, \texttt{pycausal}\footnote{\url{https://zenodo.org/record/3592985.}} (based on \texttt{tetrad}\cite{Ramsey2018}) and \texttt{dowhy-GCM} \cite{blobaum22}. 

\subsection{Generation Strategies}

We instantiate RBST with the most conservative options for the steps' alternative strategies (described in Section \ref{sec:CRBST}):

\begin{itemize}
    \item \textbf{Model building}: We opt for CSD,  with the FCI algorithm \cite{Spirtes21}, 
    one of the simplest solutions for CSD. 
    The initial database has 100 randomly generated tests;
    \item \textbf{Intervention variable selection}: We randomly select the variable for the intervention;
    \item \textbf{Intervention value assignment}: We exhaustively evaluate every possible value of the selected (categorical) variable and choose the value minimizing the test objective;
    \item \textbf{Effect estimation}: We use the simulation-based approach, producing a hypothetical test scenario for each possible value of the selected variable.\footnote{Sample size = 1,000, the \texttt{dowhy-GCM} default value.} 
\end{itemize}

The initial 
model is updated with the executed tests at each iteration. %
We compare RBST with: 
\begin{itemize}
\item[i)] an ML-driven search-based technique (SBST-ML) that, similarly to \cite{Haq22}, uses surrogate ML models to save tests execution coupled with a genetic algorithm to evolve the population of tests (same initial database as RBST); 
\item[ii)] Adaptive Random Testing (ART) \cite{Chen10}.
\end{itemize}

\subsection{Preliminary Findings}
For each technique, we run the testing session with a fixed time budget of 120 minutes, and repeated 20 times. 
The evaluation compares the technique as for  \textit{effectiveness} (number of violations found in a whole testing session) and \textit{efficiency} (number of violations at time 20-minutes time intervals).  In addition, we investigate to what extent the techniques push the output value to the ``edge" (namely, close to the threshold), and the diversity of the generated test suites.

\begin{figure}[t]
\centering
    \includegraphics[width=0.91\columnwidth]{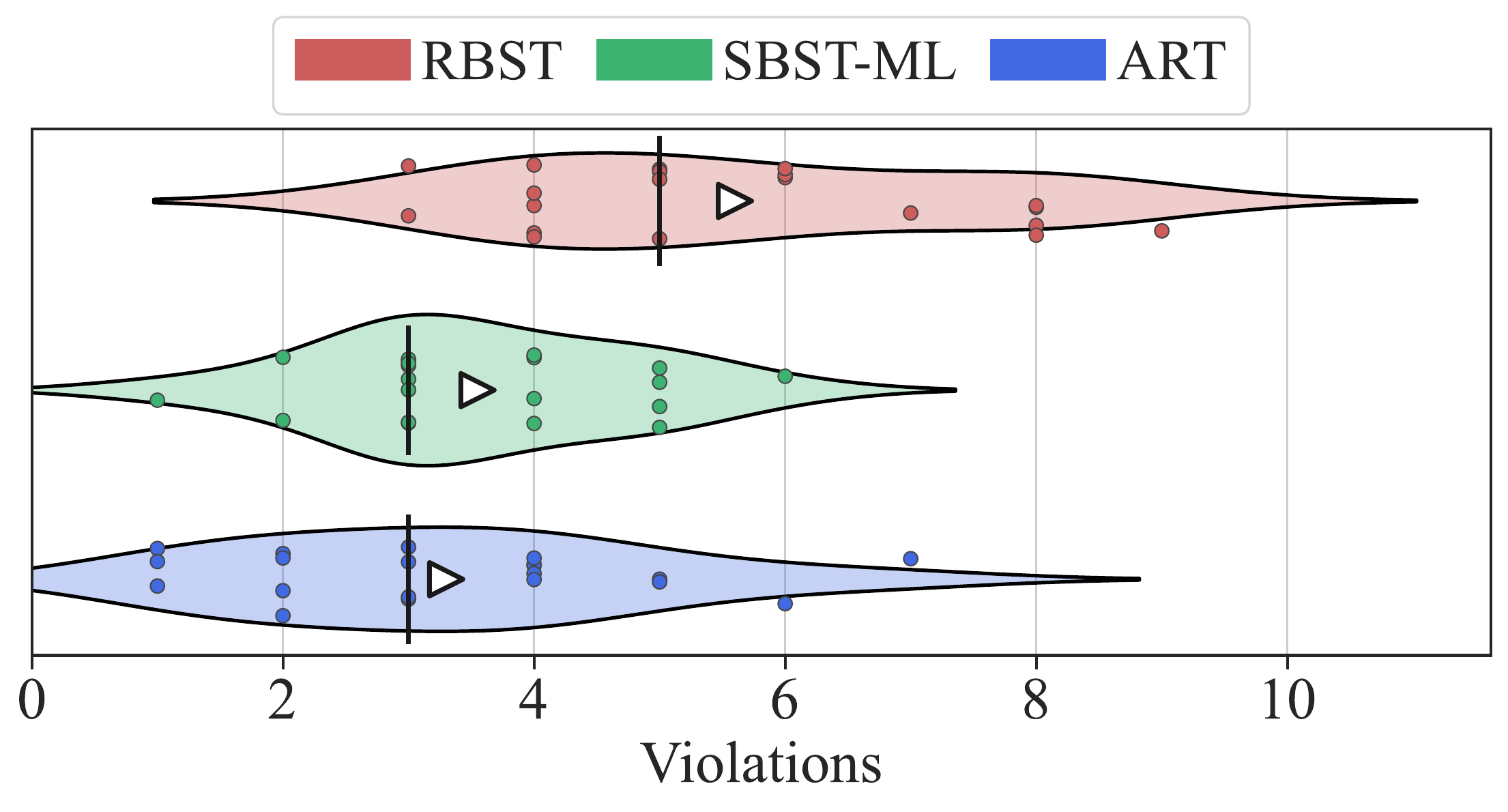}
    \caption{Effectiveness of RBST compared to SBST-ML and ART: Violin plots of the number of safety violations over test repetitions}
    \label{fig:cfails} 
\vspace{-9pt}
\end{figure}

Figure \ref{fig:cfails} shows violin plots of the number of violations found in the 20 repetitions by the three compared techniques (RBST, SBST-ML, ART). 
For every technique, the vertical black line represents the median, and the small triangle represents the mean value. RBST finds significantly more violations than the compared approaches. (The results of statistical tests are: Friedman test  \textit{p-value} 1.37E-03; pairwise Dunn test \textit{p-value}s: 1.43E-02 for RBST \textit{vs} SBST-ML;  2.20E-03 for RBST \textit{vs} ART.) 

\begin{figure}[b!]
\vspace{-9pt}
	\centering
    \includegraphics[width=0.91\columnwidth]{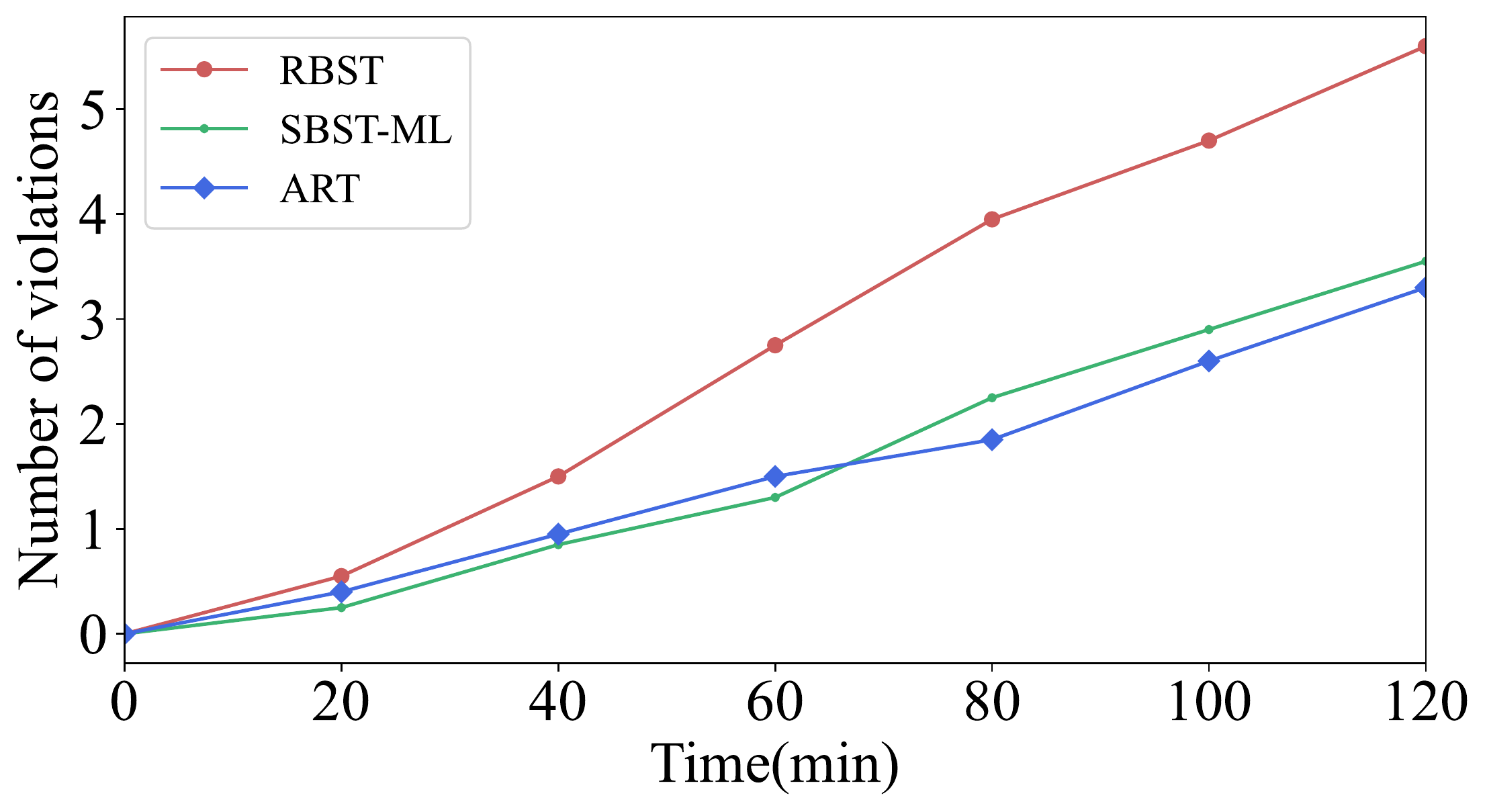}
    \caption{Efficiency of RBST compared to SBST-ML and ART: Mean number of violations over testing time}
    \label{fig:cfails_time} 
    \vspace{-9pt}
\end{figure}

Figure \ref{fig:cfails_time} shows the mean number of violations found by the three techniques over time. Since the execution of a scenario takes, on average, less than 10 minutes, we split the time budget into 20 minutes intervals to have at least 2 scenarios per slot. One can see that RBST significantly outperforms the other techniques, almost reaching the maximum of the search-based technique with half of the budget.\\

\begin{figure}[t]
	\centering
    \includegraphics[width=0.99\columnwidth]{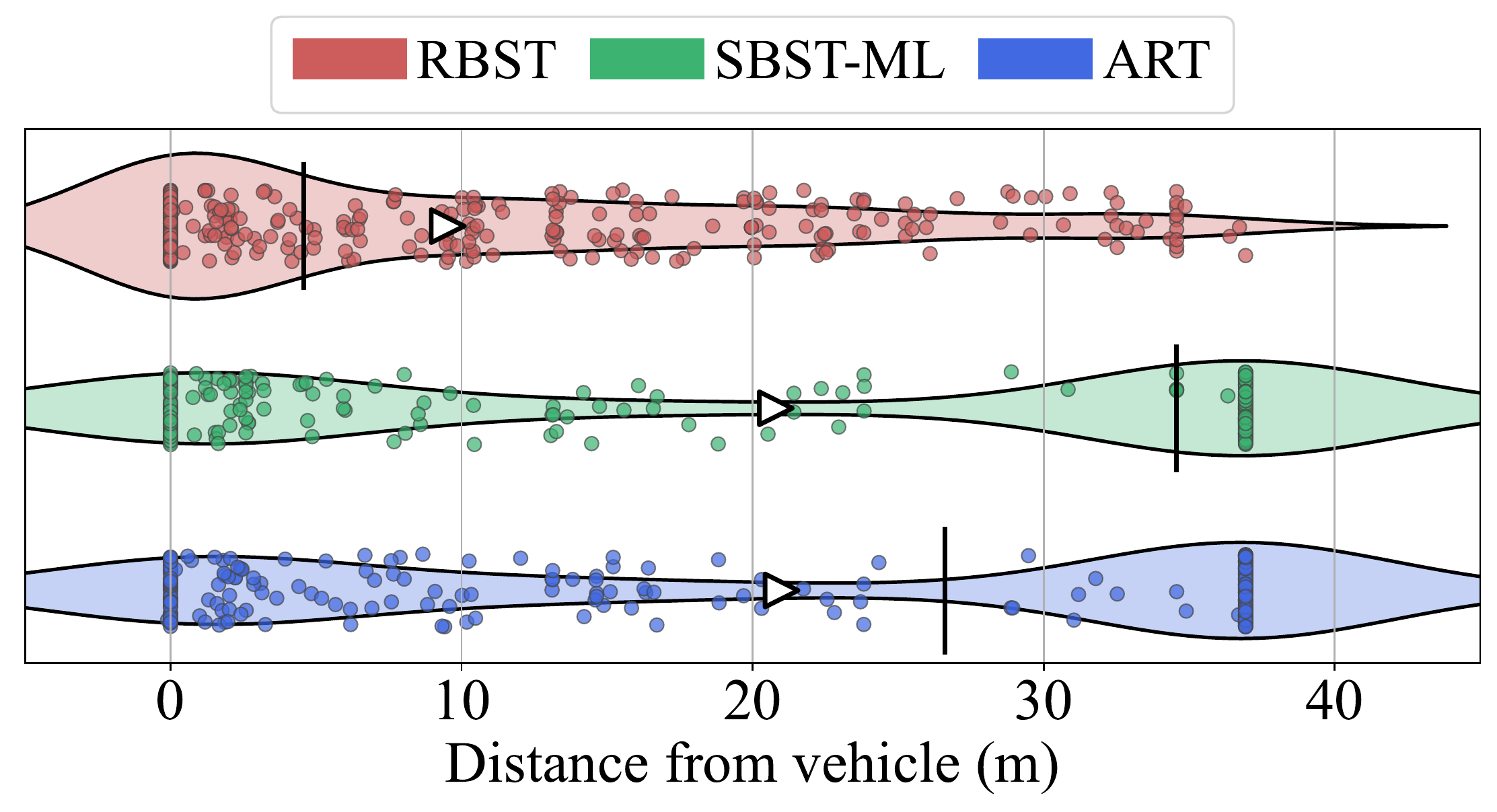}
    \caption{Violing plots of the distance from other vehicles for the (values close to zero correspond to near-violating scenarios)}
    \label{fig:zscore_box} 
    \vspace{-6pt}
\end{figure}

To investigate near-violating scenarios, we collect the output value of each scenario execution. As the output corresponds to the minimum distance from other vehicles, values close to zero represent test scenarios close to violating a safety requirement. 
Figure \ref{fig:zscore_box} shows violin plots of the results for the three compared techniques. RBST achieves the best values (Friedman \textit{p-value} 3.23E-13; Dunn \textit{p-value}s: $<$1.00E-04 \textit{vs} both SBST-ML and ART), while SBST-ML and ART do not differ (Dunn \textit{p-value} = 1.00E+00). 

\begin{figure}[b]
	\centering
    \includegraphics[width=0.99\columnwidth]{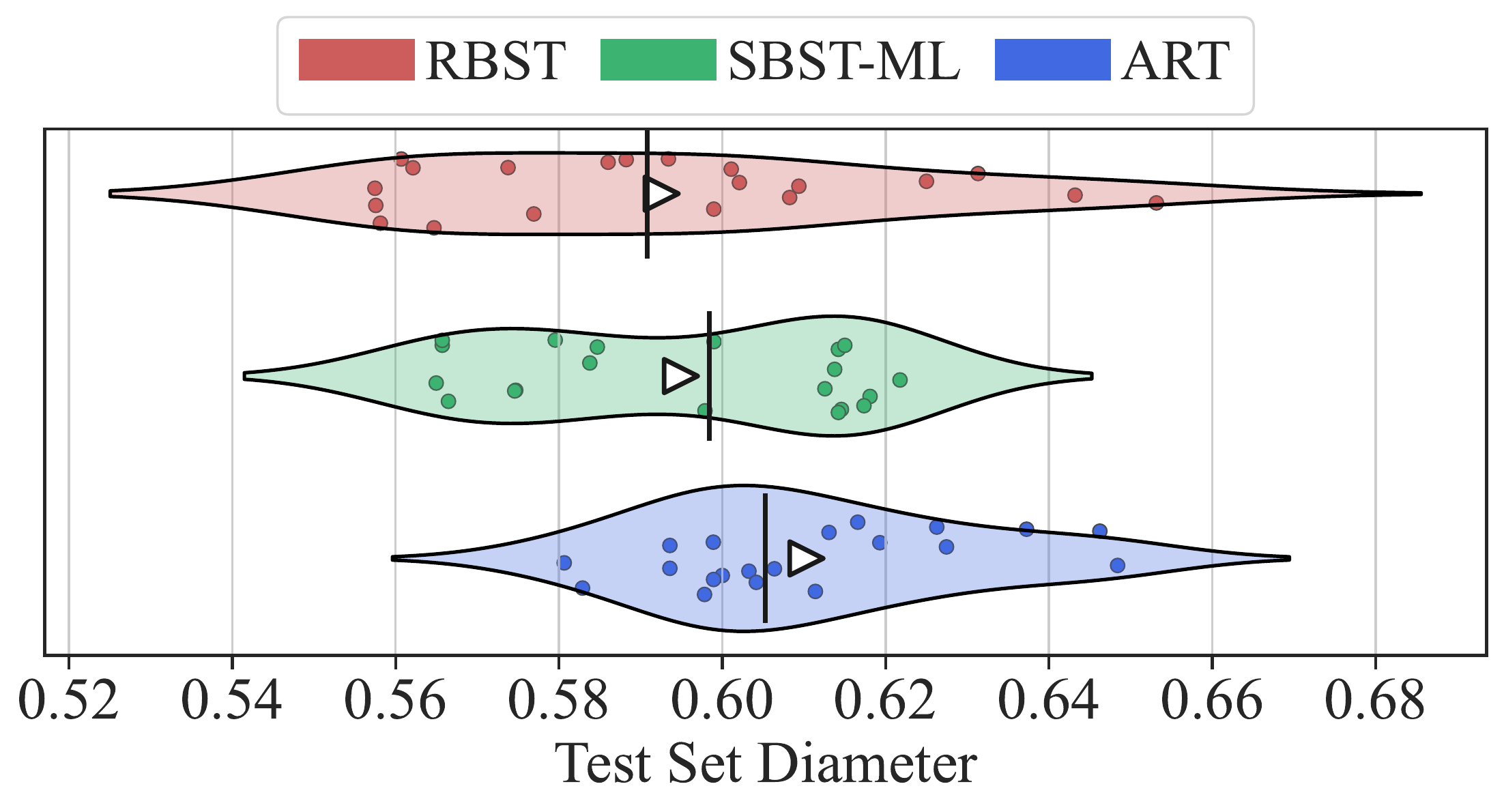}
    \caption{Comparison of techniques as for diversity of the test suites: Violin plots of the Test Set Diameter on input over test repetitions}
    \label{fig:diversity} 
\end{figure}

Figure \ref{fig:diversity} reports on the diversity of the test suites, estimated with the Test Set Diameter on input (TSD-I), a well-known metric for black-box test suite diversity \cite{Feldt16,Miranda18,Henard16}. RBST achieves the worst values of median and mean, but with  %
no significant difference 
 (Friedman \textit{p-value}: 2.93E-01). %

Due to the causal inference step, RBST entails a time overhead. Generating a test case required, on average, $t_{gen}$=$1.44$ seconds, which is negligible with respect to the average test execution time ($t_{gen}$ = 0.34\% $t_{exec}$). 

\label{sec:experiments}

\section{Future Plans}
Reasoning-Based Software Testing is conceived to foster the use of causal reasoning in software testing. Depending on the testing task (e.g., test generation, test prioritization), on the objective (e.g., fault detection, coverage), and especially on how the RBST process steps are instantiated (Table \ref{tab:operators}), 
several techniques can be implemented in the next years.  

Our short-term plan is to explore RBST alternative instantiations for tests generation, so as to give criteria for their best implementation. For instance, performance could be considerably improved by choosing the right trade-off between enhancing the model (such as uncertainty minimization) and attaining the testing aim. 

Strategies for exploring the space of possible interventions are also a big opportunity for improvement. Then, interpretability/explainability of causal models will be investigated, as this allows the seamless integration of human knowledge.

In the medium-long term, we plan to target regression testing and to explore multiple testing objectives.  
Moreover, we aim at integrating counterfactual reasoning in the loop for post-test execution analysis, so as to support debugging via 
\textit{actual causation} analysis (i.e., the assignment of causal responsibility for an occurred event). 
Lastly, we plan to broaden the RBST application to a variety of domains to enforce external validity.  %

\label{sec:future}

\section*{Acknowledgement}
This project has received funding from the European Union’s Horizon 2020 research and innovation programme under the Marie Sk{\l}odowska-Curie grant agreement No 871342. It is also supported by the DIETI COSMIC project.

\bibliographystyle{IEEEtran}
\IEEEtriggeratref{25}

\end{document}